\documentclass[superscriptaddress,aps,journal=prl,preprint]{revtex4-2}
\usepackage{CJK}  % For Chinese characters

\usepackage{graphicx}
\usepackage{epstopdf}
\usepackage{bm,amsmath,amssymb} % for math
\usepackage{color, soul} % for highlight
\usepackage{dcolumn}   % needed for some tables
\usepackage{multirow}
\usepackage{makecell}
\usepackage{threeparttable}
\usepackage{verbatim}
\usepackage{tabularx}
\newcommand{\etal}{{\em et al.\ }}

\newcommand{\PTOPSTO}{(PTO)$_{10}$/(PSTO)$_{10}$}
\newcommand{\PTOSTO}{(PTO)$_{10}$/(STO)$_{10}$}
\newcommand{\PTOSTOall}{(PbTiO$_3$)$_{10}$/(SrTiO$_3$)$_{10}$}
\newcommand{\PTOPSTOall}{(PbTiO$_3$)$_{10}$/(Pb$_{0.4}$Sr$_{0.6}$TiO$_3$)$_{10}$}
\begin{document}
\begin{CJK*}{UTF8}{gbsn}
% Start the CJK environment with UTF-8 encoding and simplified Chinese font

\title{Topological phase transitions in perovskite superlattices driven by temperature, electric field, and doping}

\author{Jiyuan Yang}
%\author{Jiyuan Yang}
\affiliation{Zhejiang University, Hangzhou, Zhejiang 310058, China}
\affiliation{Department of Physics, School of Science and Research Center for Industries of the Future, Westlake University, Hangzhou, Zhejiang 310030, China}
\author{Shi Liu}
%\author{Shi Liu}
\email{liushi@westlake.edu.cn}
\affiliation{Department of Physics, School of Science and Research Center for Industries of the Future, Westlake University, Hangzhou, Zhejiang 310030, China}
\affiliation{Institute of Natural Sciences, Westlake Institute for Advanced Study, Hangzhou, Zhejiang 310024, China}

%\date{\today}

\newpage
\begin{abstract}{ 
Many dipolar topological structures have been experimentally demonstrated in (PbTiO$_3$)$_n$/(SrTiO$_3$)$_n$ superlattices, such as flux-closure, vortice, and skyrmion. In this work, we employ deep potential molecular dynamics (MD) to  investigate the atomic-level dynamical response of the \PTOSTOall~superlattice, which hosts polar vortex arrays, to variations in temperature and electric field. Our simulations reveal a unique phase transition sequence from ferroelectric-like to antiferroelectric-like to paraelectric in the in-plane direction as temperature increases.
In the ferroelectric-like state, we observe field-driven reversible switching of in-plane polarization coupled with out-of-plane movements of vortex cores. 
In the antiferroelectric-like region, the polarization-electric field hysteresis loop exhibits a superparaelectric feature, showing nearly no loss. This behavior is attributed to a strong recovering force to form polar vortex arrays, dictated by the electrical and mechanical boundary conditions within the superlattice. The \PTOSTOall~superlattice in the antiferroelectric-like state also demonstrates large in-plane susceptibility and tunability. The effect of Pb doping in the SrTiO$_3$ layer on the topological structural transition in the superlattice is investigated. The weakened depolarization field in the PbTiO$_3$ layers leads to new dipolar configurations, such as enlarged skyrmion bubble within $c$ domains in \PTOPSTOall, and we quantify their thermal and electrical responses using MD simulations. These quantitative atomistic insights will be useful for the controlled optimization of perovskite superlattices for various device applications.
}
\end{abstract}

\maketitle
\end{CJK*}

\newpage
\section{Introduction}
Ferroelectric materials, known for their spontaneous polarization that can be reversed by an external electric field, have been a focus of research due to their unique properties and diverse applications. Ferroelectrics have been utilized in non-volatile memory devices~\cite{Guo13p1990,Garcia14p4289}, sensors~\cite{Damjanovic01p191}, and actuators~\cite{Park97p20}, capitalizing on their ability to retain polarization states in the absence of external bias and their strong coupling of the polarization to external stimuli. Over the past decade, considerable progress has been made in understanding and controlling the spontaneously formed topological polar structures in ferroelectric oxide heterostructures, including flux-closure~\cite{Tang15p547},vortex~\cite{Yadav16p198}, skyrmion~\cite{Das19p368}, meron~\cite{Wang20p881}, and dipole wave~\cite{Lu18p177601,Gong21p5503}. These nontrivial ferroelectric topological structures, often regarded as the electric counterparts of their spin analogues, are formed due to the delicate competitions between electric, gradient and elastic energy terms. 

Previously, developed theoretical tools, such as second-principles calculations and phase-field simulations, have  been highly successful in advancing the understanding and designing ferroelectric topological structures~\cite{Yadav16p198,Yamada07p270,Das19p368,Wang20p881,Lu18p177601,Gong21p5503,Hong17p2246,Junquera23p025001,Liu22p34,Prokhorenko24p412,Behera22peabj8030,Behera23p2208367}. More recently, Yuan \etal~revealed the
existence of a field-induced hexagonal-lattice skyrmion crystal in PbTiO$_3$ thin films through phase-field simulations~\cite{Yuan23p226801}. Aramberri \etal~employed a second-principles
approach to investigate the dynamics of electric skyrmions, demonstrating the quasiparticle nature of these electric bubbles~\cite{Aramberri24p136801}.
The main methods to control the dipolar topology  include strain engineering and periodicity modulation. Taking (PbTiO$_3$)$_n$/(SrTiO$_3$)$_n$ [(PTO)$_n$/(STO)$_n$] superlattice as an example, in-plane strain arising from the lattice mismatch between the superlattice and the substrate can influence the types of topological structures that emerge. For instance, vortex-type topological structures form under tensile strain~\cite{Yadav16p198}, while dipolar skyrmions appear under slightly compressive strain~\cite{Das19p368}. The periodicity of the superlattice, defined by the number of PTO and STO layers, also has a significant impact on the formation of topological structures. For example, Hong \etal. revealed through phase-field simulations that in (PTO)$_n$/(STO)$_n$, as $n$ increases, the domain structure evolves from $a_1/a_2$ to vortex and eventually to flux-closure~\cite{Hong17p2246}. However, the impact of doping on the formation, stability, and dynamical properties of topological structures in ferroelectric superlattices remains largely unexplored at the atomic level.

In this work, we employ a machine-learning-based deep potential (DP) model and perform molecular dynamics (MD) simulations to investigate the temperature- and electric-field-driven evolution of polar vortex arrays in (PTO)$_{10}$/(STO)$_{10}$ superlattices. Our results reveal a novel temperature-driven phase transition sequence in the in-plane direction, where the system transforms from a ferroelectric-like (FE$^*$) to an antiferroelectric-like (AFE$^*$) and finally to a paraelectric (PE) state, distinct from the conventional ferroelectric-paraelectric (FE-PE) or antiferroelectric-paraelectric (AFE-PE) phase transition observed in single crystals~\cite{Shirane51p265,Roleder00p287,Haas65pA863,Grunebohm21p073002,Grinberg02p909}. The counterintuitive temperature-driven FE$^*$-to-AFE$^*$ transition highlights the potential of harnessing topological domain structures to engineer properties that are inaccessible in single-domain states. 
In the AFE$^*$ region, the polarization-electric field loop of the \PTOSTO~superlattice is nearly hysteresis-free, making it ideal for low-loss applications. We find that the introduction of Pb dopants into the STO layers can modulate the depolarization field in the PTO layers, leading to the emergence of new topological structures in (PbTiO$_3$)$_{10}$/(Pb$_{\alpha}$Sr$_{1-\alpha}$TiO$_3$)$_{10}$ [\PTOPSTO] superlattices, such as enlarged skyrmion bubbles embedded in $c$ domains. 
Finally, we show that the combination of strain, periodicity, and doping engineering allows for controlled design of a rich spectrum of polar topological patterns with distinct properties. This multi-factor approach not only provides a versatile platform for various applications, including high-density data storage, low-power memory devices, and advanced sensors, but also opens new avenues for exploring the fundamental physics of topological structures in ferroelectric materials. 

\section{Computational Methods}
\subsection{Deep potential of PSTO}
The essence of the deep potential (DP) model is to construct a deep neural network that maps local atomic environments to atomic energies and forces, as extensively discussed in previous studies~\cite{Zhang18p143001,Zhang18p4436,Zhang20p107206}. To train our DP model for PSTO solid solutions, we built a dataset containing the DFT energies and forces of various configurations of PTO, STO, and Pb$_\alpha$Sr$_{1-\alpha}$TiO$_3$. The dataset includes 13,021 PTO structures in the $P4mm$ and $Pm\Bar{3}m$ phases, 3,538 STO structures in the $I4/mcm$ and $Pm\Bar{3}m$ phases, and 2,560 PSTO structures with compositions of $\alpha=0.25$, 0.5, and 0.75. All DFT calculations are performed with the Vienna Ab initio Simulation (VASP) package~\cite{Kresse96p11169, Kresse96p15} using the projected augmented wave (PAW)
method~\cite{Blochi94p17953,Kresse99p1758}. Perdew-Burke-Ernzerhof functional modified for solids (PBEsol)~\cite{Perdew08p136406} within the generalized gradient approximation is chosen as the exchange-correlation functional. An energy cutoff of 800 eV and a $k$-point spacing of 0.3~\AA$^{-1}$ are sufficient to converge the energy and atomic forces. Additional details on model development can be found in ref.~\cite{Wu23p144102}. Density functional perturbation theory (DFPT) is employed to directly calculate the first-order response of the wavefunction (charge density) to an applied electric field for computing the Born effective charges~\cite{Gajdovs06p045112,Baroni86p7017}. The energy cutoff and $k$-point spacing settings in the DFPT calculations are kept the same as those used in constructing the training dataset. 
This DP model accurately describes the phase diagram of PSTO across the entire temperature and composition range, and successfully reproduce the experimentally observed strain-induced topological structure transition~\cite{Wu23p144102}. We have developed an online {\href{https://github.com/JiyuanY/L52-topological}{{notebook}}}~\cite{L52_data} on Github, which provides access to the training database, force field model, and training metadata. 

\subsection{MD simulations}
To simulate the (PTO)$_{10}$/(STO)$_{10}$ superlattice, we construct a 40$\times$20$\times$20 supercell containing 80,000 atoms. 
We note that PBEsol is chosen as the exchange-correlation functional for constructing the DFT database used to train the DP model. Since the ground-state lattice constants of typical perovskite oxides, such as PbTiO$_3$ and SrTiO$_3$, predicted by PBEsol deviate slightly from experimental values, it is expected that the theoretical mechanical boundary conditions leading to the formation of a specific topological structure will differ slightly from those observed experimentally. In this study, we focus specifically on the pure vortex state. We explored various combinations of in-plane lattice parameters, $a_{\rm IP}$ and $b_{\rm IP}$, and found that $a_\mathrm{IP}=3.937$ \AA~and $b_\mathrm{IP}=3.929$~\AA~leads to well-defined polar vortex arrays. In reference to the in-plane ground-state lattice parameter of PTO at 300~K ($a_\mathrm{PTO}=3.919$~\AA) in our MD simulations, the strain state is $\delta_x=0.46\%$ and $\delta_y=0.26\%$, and the average strain is $\delta_\mathrm{ave}=0.36\%$. The value of $\delta_\mathrm{ave}$ agrees reasonably well with the experimental value of 0.3\% that yields a pure vortex state~\cite{Das23p6602} and the 1.1\% strain that leads to a mixed state of $a_1$/$a_2$ and vortex arrays~\cite{Yadav16p198,Li21p376}.
To simulate the (PTO)$_{10}$/(PSTO)$_{10}$ superlattice, Pb atoms are randomly doped into the STO layers. MD simulations in the constant-temperature constant-pressure ($NPT$) ensemble are carried out using the LAMMPS package~\cite{Plimpton95p1}. We note that LAMMPS allows for independent specification of all six components of an external stress tensor. This capability enables us to control each of the six dimensions of the simulation box independently.
In our simulations, we fixed the in-plane lattice constants while allowing the out-of-plane lattice constant to adjust in response to the target pressure.
In all simulations, a timestep of 2 fs is used for integrating the equations of motion. The system temperature is controlled using the No\'e-Hoover thermostat, while the Parrinello-Rahman barostat is employed to maintain the pressure. During the simulation of temperature-driven phase transitions, the pressure is held at 1.0 bar. For each temperature, the system is first equilibrated for 50 ps, followed by a 50 ps production run. We find that this simulation duration is adequate to achieve convergence of the structural parameters.

We utilize the ``force method"~\cite{Umari02p157602,Liu16p360} in classical MD to simulate the effects of external electric fields on the dynamical response of topological structures. In this method, the field-induced force $F_i$ on ion $i$ is the product of its Born effective charge (BEC) tensor $Z_i^*$ and the electric field strength $E$. To obtain the BEC tensors, we employ the DFPT method~\cite{Gajdovs06p045112,Baroni86p7017} to calculate the values for ions in the unit cells of the PTO tetragonal phase and the STO cubic phase, as reported in Table~\ref{BECs}. The polarization of a unit cell at time $t$ can be expressed as:
\begin{equation}
P^m(t)=\frac{1}{V_{\rm uc}}\left[\frac{1}{8} Z_{A}^* \sum_{k=1}^8 \mathbf{r}_{A, k}^m(t)+Z_{\mathrm{Ti}}^* \mathbf{r}_{\mathrm{Ti}}^m(t)+\frac{1}{2} Z_{\mathrm{O}}^* \sum_{k=1}^6 \mathbf{r}_{\mathrm{O}, k}^m(t)\right]
\end{equation}
where $P^m(t)$ is the polarization of unit cell $m$ at time $t$; $V_{\rm uc}$ is the volume of the unit cell; $Z_{A}^*$, $Z_{\mathrm{Ti}}^*$, and $Z_{\mathrm{O}}^*$ are the BECs of $A$-site (Pb or Sr), Ti, and O atoms, respectively; $\mathbf{r}_{A, k}^m(t)$, $\mathbf{r}_{\mathrm{Ti}}^m(t)$, and $\mathbf{r}_{\mathrm{O}, k}^m(t)$ are the atomic coordinates in unit cell $m$ at time $t$. Here, the local polarization is approximated as the electric dipole moment per unit cell. 
In the treatment of Pb$_\alpha$Sr$_{1-\alpha}$TiO$_3$ solid solutions, the BECs of $A$-site ions are assigned with the averaged value as  $Z_{A}^*=\alpha Z_{\mathrm{Pb}}^*+(1-\alpha)Z_{\mathrm{Sr}}^*$, with the charges of Ti and O atoms determined in a similar manner, as Ti and O ions each have different BECs in PTO and STO. This ensures local charge neutrality to properly define a local electric dipole.

\section{Results and Discussions}
\subsection{Temperature-driven phase transitions in superlattices}
\begin{figure}[]
\centering
\includegraphics[width=1.0\textwidth]{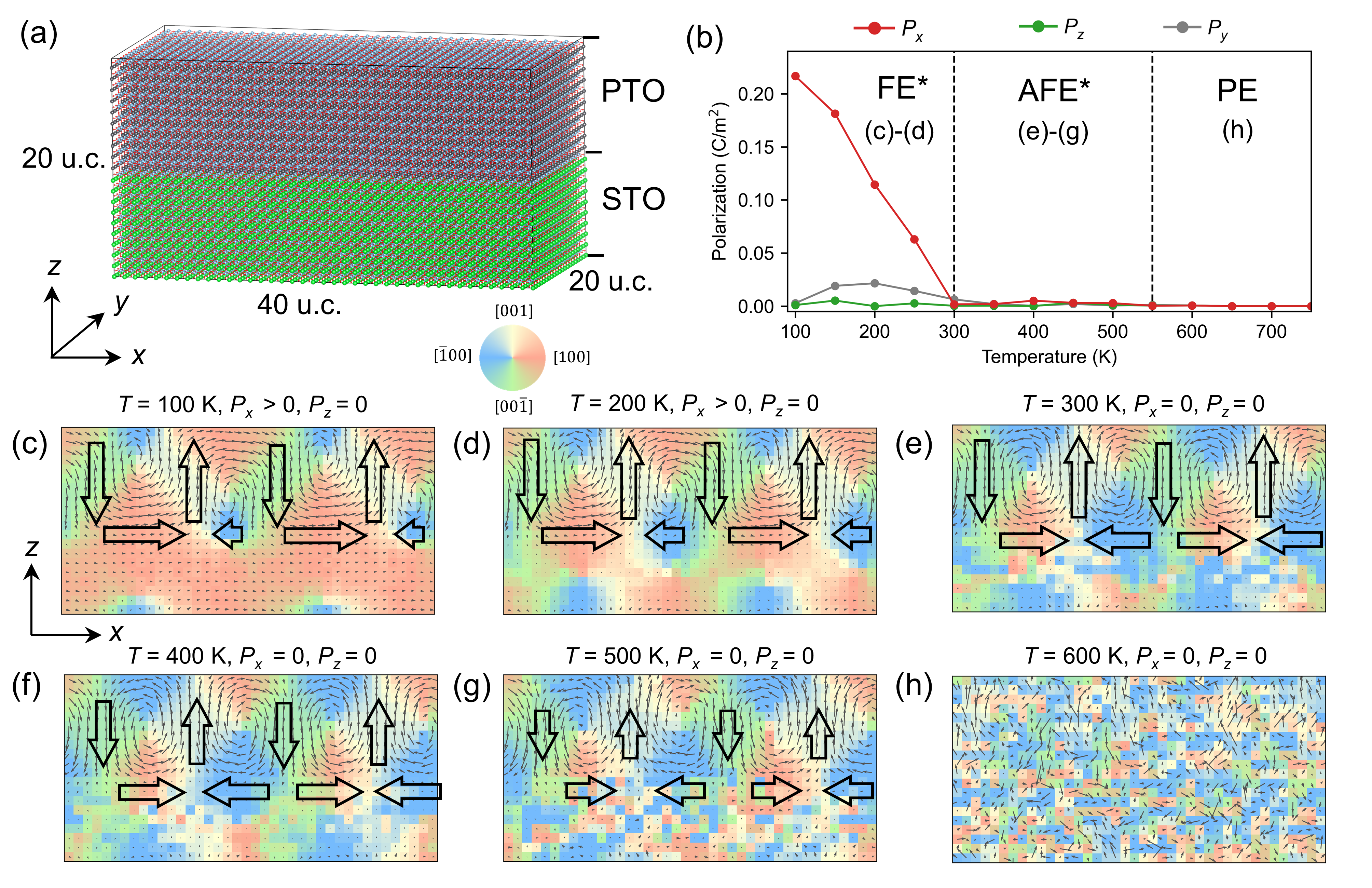}
\caption{(a) A 40$\times$20$\times$20 supercell used to model the (PTO)$_{10}$/(STO)$_{10}$ superlattice with periodic boundary conditions in MD simulations. (b) Temperature dependence of the average polarization of the entire superlattice along the $x$, $y$ and $z$ axes. Snapshots of the $xz$ plane (c)-(h) show representative states at different temperatures, including (c) the ferroelectric-like (FE$^*$) state at 100~K, (e) the antiferroelectric-like (AFE$^*$) state at 300~K, and (h) the paraelectric (PE) state at 600~K. Each small black arrow denotes the local polarization calculated using a 5-atom unit cell, while the large empty black arrows represent the polarization directions of nanodomains. The background is colored based on the polarization direction. 
}
\label{PhaseTrans}
\end{figure}

We investigate the temperature-dependent structural evolution in the prototypical \PTOSTO~superlattice that supports polar vortex arrays, as shown in Fig.~\ref{PhaseTrans}a. 
The in-plane ($P_x$ and $P_y$) and out-of-plane polarization ($P_z$) of the superlattice are analyzed as functions of temperature ($T$). It is found that the out-of-plane polarization of the entire system remains zero over the entire temperature range from 100 K to 700~K. In contrast, the in-plane $P_x$ polarization transitions from non-zero to zero at 300~K, as depicted in Fig.~\ref{PhaseTrans}b. 

\begin{figure}[]
\centering
\includegraphics[width=1.0\textwidth]{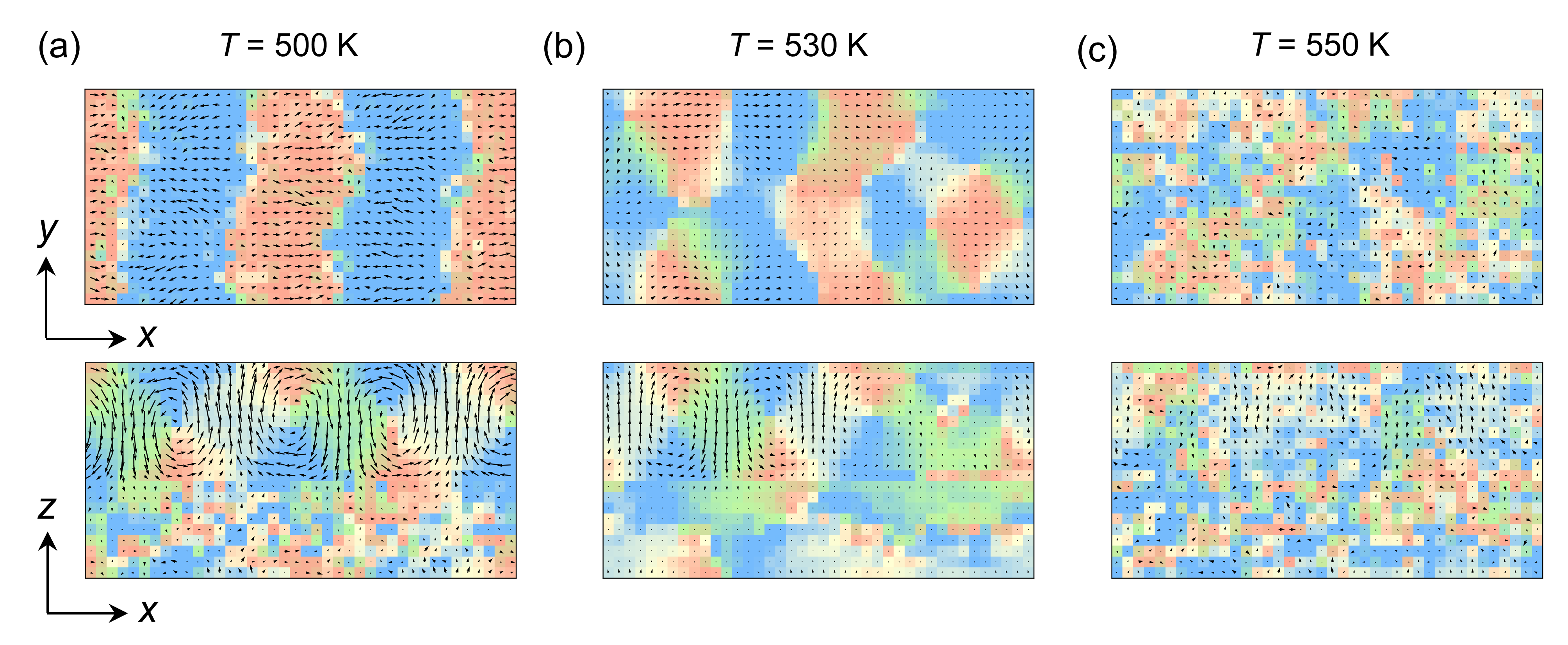}
\caption{Snapshots of topological structures at temperatures near the transition from AFE$^*$ to PE. The superlattice transitions from (a) a vortex array state to (b) a polar liquid state, and finally to (c) a paraelectric phase.}
\label{polar_liquid}
\end{figure}

We examine the microscopic domain structures at several representative temperatures using MD simulations. At a low temperature of 100~K, the vortex cores are situated near the PTO/STO interfaces, forming a zigzag chain.
The arrangement of alternating $+P_z$ and $-P_z$ domains, each approximately 5 unit cells wide, results in a cancellation of out-of-plane polarization. Interestingly, the zigzag vortex cores break global in-plane inversion symmetry, leading to alternating $+P_x$ and $-P_x$ domains of unequal size. This asymmetry prevents the polarization of $+P_x$ and $-P_x$ domains from completely canceling out, giving rise to a ferroelectric-like (FE$^*$) state with net polarization along the $x$ axis (Fig.~\ref{PhaseTrans}c). 
We note that it is more rigorous to label this domain structure featuring unequal volume fractions of $+P_x$ and $-P_x$ domains as ``ferrielectric-like". For convenience, we will continue to use the term ``ferroelectric-like" in the following discussion.
As the temperature increases, we observe a collective shift of vortex cores away from the interfaces (Fig.~\ref{PhaseTrans}d).  At 300~K as shown in Fig.~\ref{PhaseTrans}e, vortex cores are aligned along the $x$ axis, leading to equal volumes of $P_x$ and $-P_x$ domains. This configuration, termed antiferroelectric-like (AFE$^*$), features zero net in-plane polarization. The AFE$^*$ state remains stable up to 550~K (Fig.~\ref{PhaseTrans}f-g).  At a higher temperature of 600~K, the entire superlattice transitions to a paraelectric state, characterized by randomly oriented local dipoles (Fig.~\ref{PhaseTrans}h). The conterintuitive temperature-driven FE$^*$-AFE$^*$ transition in the \PTOSTO~superlattice is unusual and distinct from the behavior of single-crystal ferroelectrics, which rarely evolve into an antiferroelectric state with increasing temperature. Additionally, a careful examination of the polar structures around the AFE$^*$-PE transition temperature reveals the presence of an intermediate topological phase. As shown in Fig.~\ref{polar_liquid}, the transition is not sharp, mediated by a complex polar liquid phase characterized by the loss of long-range translational order along the $y$ axis. This observation aligns with the recently reported two-step melting process, which involves an intermediate hexatic-like phase~\cite{Gomez24p066801,Gomez22pL220103}.

\begin{figure}[]
\centering
\includegraphics[width=1.0\textwidth]{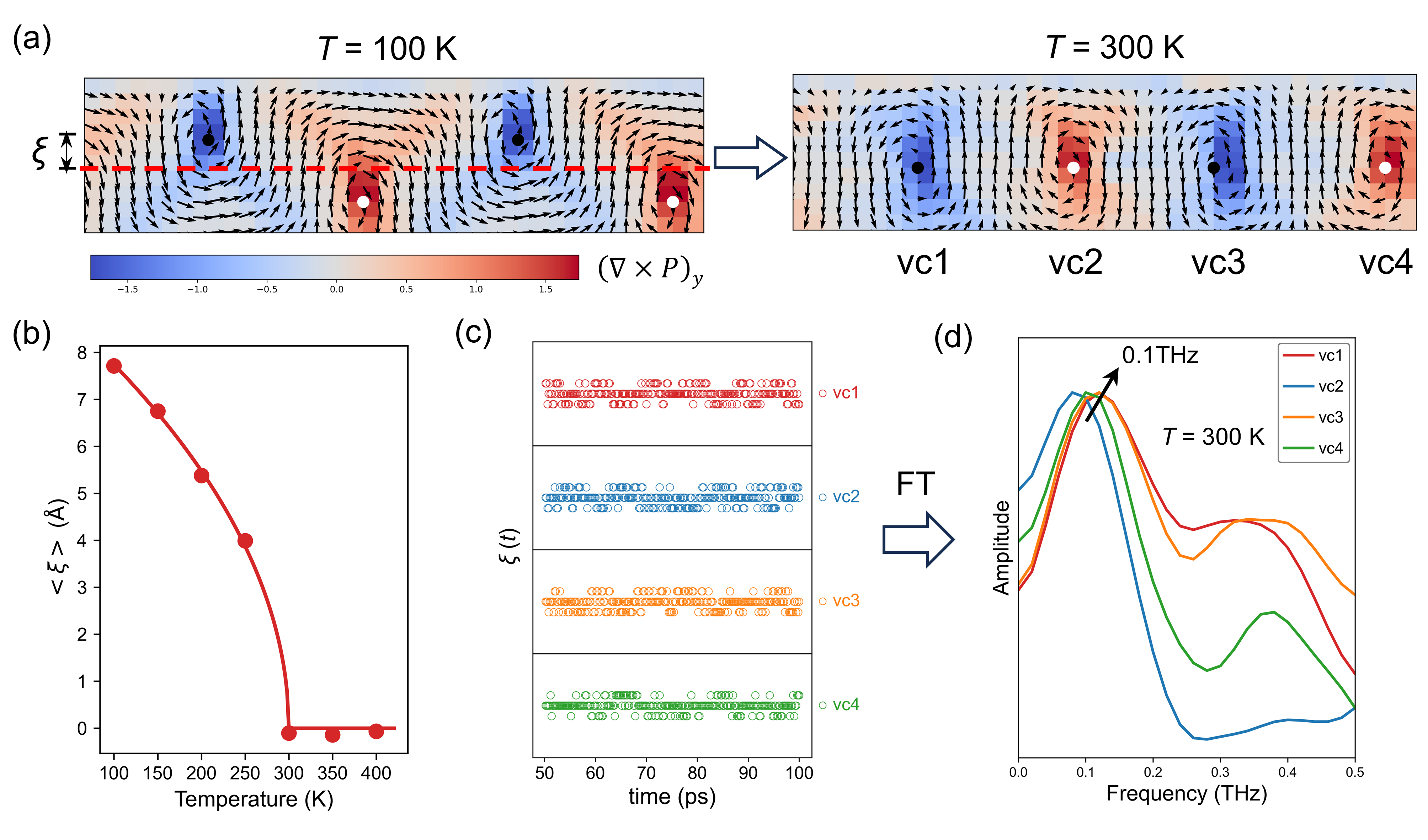}
\caption{(a) Definition of the order parameter $\xi$, representing the out-of-plane distance from the vortex core to the center of the PbTiO$_3$ layer. (b) Temperature dependence of the ensemble-averaged $\xi$ obtained from MD simulations, with the data fitted to $\left<\xi\right> \propto \sqrt{300-T}$, shown as the solid line. (c) Time evolution of $\xi$ for the four vortex cores (vc) depicted in (a). (d) Fourier transform (FT) of $\xi(t)$.}
\label{vortexon}
\end{figure}

The collective movement of vortex cores is a manifestation of the vortexon mode~\cite{Li21p376,Yang21pL220303}.
We track the position of the vortex core in MD simulations by identifying the extremal points of vorticity, as illustrated in Fig.~\ref{vortexon}a. We introduced a new order parameter $\xi$, defined as the out-of-plane distance between the vortex core to the center of the PbTiO$_3$ layer. Figure ~\ref{vortexon}b presents the temperature dependence of the ensemble-averaged value of $\xi$ from MD simulations, revealing a displacive, Landau-type first-order phase transition. Furthermore, we perform a Fourier transform on the time evolution of $\xi$ at 300~K (Fig.~\ref{vortexon}c-d) and obtain the frequency spectrum,  which shows a peak at 0.1~THz, closely aligning with the experimental value of 0.08 THz~\cite{Li21p376}. These results confirm the presence of vortexon modes and validate the accuracy of the DP model utilized in this study. Moreover, the displacement of the vortex core affects the ratio of $+P_x$ domains and $-P_x$ domains, thereby influencing the magnitude of the total $P_x$. Additionally, the (local) polarization of PbTiO$_3$ also scales with $\sqrt{T_c-T}$, where $T_c$ is the intrinsic Curie temperature. The superposition of these two effects make $P_x$ decreases more rapidly, as shown in Fig.~\ref{PhaseTrans}b.

We note that the collective shift of vortex cores can also be achieved by tuning the in-plane strain. %as demonstrated in previous studies~\cite{}. 
Specifically, Gong \etal~demonstrated that as the in-plane strain increased, the pattern in a \PTOSTO~superlattice evolved from aligned polar vortex lattice to shifted vortex lattice, and finally to an electric dipole wave~\cite{Gong21p5503}.
However, in practical applications, dynamical temperature control is often a more feasible and accessible means of manipulation. Utilizing existing thermal management technologies, such as heating elements or Peltier devices~\cite{Mannella14p234,Huang00p197}, could enable precise and rapid temperature adjustments, allowing for the stabilization of the superlattice at desired temperatures. This ease of temperature control, compared to the complexities of mechanical strain application, can potentially facilitate more robust on-demand control of polar topologies. 

\subsection{Temperature-dependent polarization-electric field hysteresis loops}
\begin{figure}[]
\centering
\includegraphics[width=1.0\textwidth]{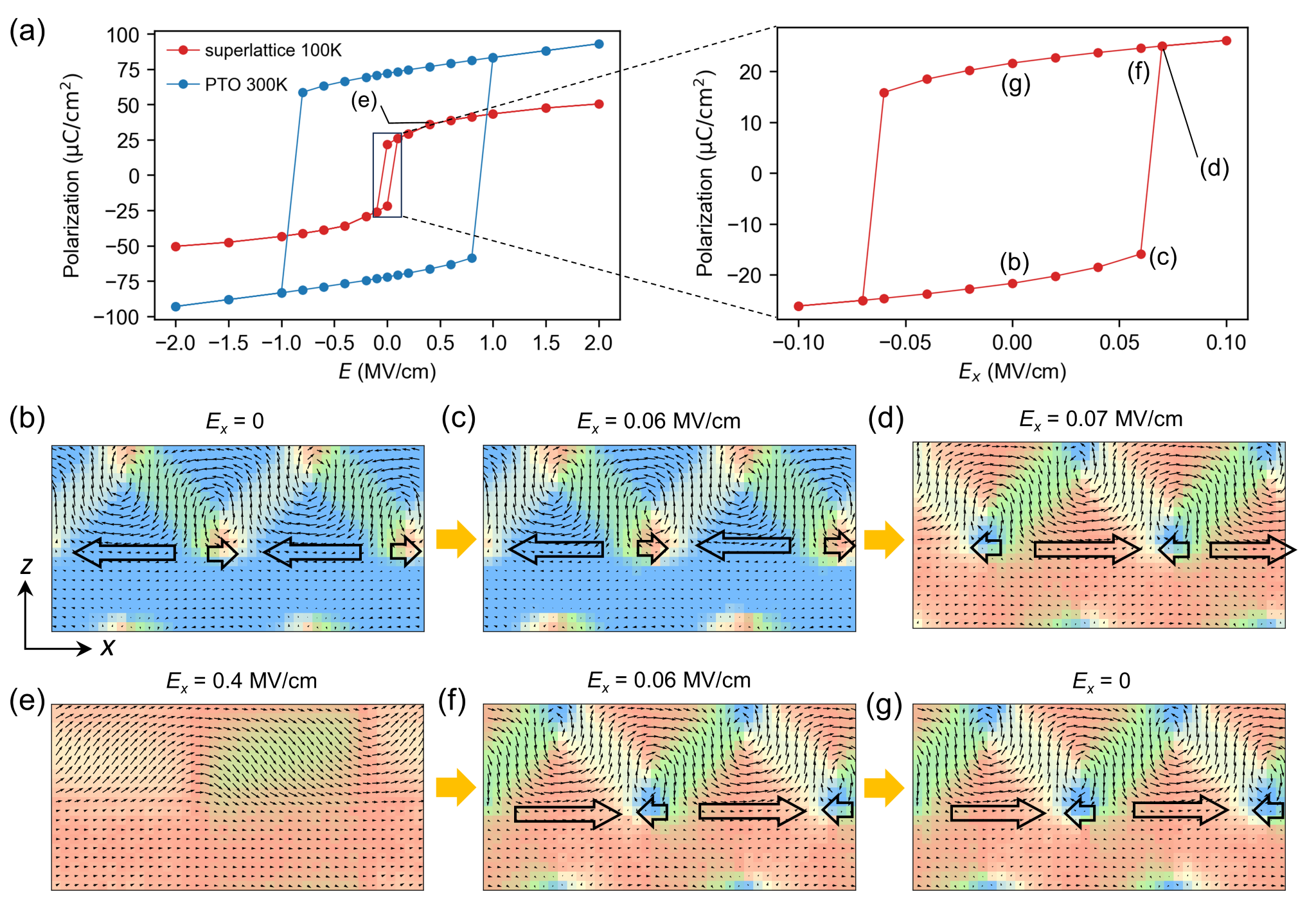}
\caption{(a) Polarization-electric field ($P_x$-$E_x$) hysteresis loop of the FE$^*$ state featuring shifted vortex cores in the \PTOSTO~superlattice at 100 K (red loop), compared to the hysteresis loop of a single-crystal single-domain PbTiO$_3$ at 300 K (blue loop). The right figure shows a magnified view of the loop for $E_x$ in the range between $-0.1$ MV/cm and 0.1 MV/cm. 
Snapshots of topological structures throughout the loop, obtained from MD simulations, are presented in (b)-(g). The color scheme is the same as that in Fig.~1.
}
\label{PE_loop_100K}
\end{figure}

The emergence of FE$^*$ and AFE$^*$ states along the in-plane direction of the polar vortex arrays raises an interesting question: do these states  exhibit polarization-electric hysteresis loops similar to those of conventional ferroelectric and antiferroelectric materials? To explore this, we simulate the $P_x$-$E_x$ loop by applying an electric field along the $x$ axis (see the red loop in Fig.~\ref{PE_loop_100K}a) and track the evolution of topological structures throughout the loop. For comparison, we also simulate the $P$-$E$ loop for a single-crystal single-domain PTO (see the blue loop in Fig.~\ref{PE_loop_100K}a)

At a low temperature of $T=100$~K, the initial state of the superlattice exhibits shifted vortex cores associated with a net $-P_x$ polarization (Fig.~\ref{PE_loop_100K}b). Applying an electric field in the $+x$ direction increases the volume of $+P_x$ domains (Fig.~\ref{PE_loop_100K}c), accompanied by a vertical shift of vortex cores. 
A sharp transition in the domain structure occurs at $E_x=0.07$~MV/cm, where the cores have completely traversed the entire PTO layers vertically and the volume of $+P_x$ domains surpasses that of $-P_x$ domains (Fig.~\ref{PE_loop_100K}d). Further increasing the field strength drives the superlattice into a dipole-wave state (Fig.~\ref{PE_loop_100K}e) and ultimately into a single $a$-domain state in the PTO layers. Following the reduction of the field strength, the superlattices returns to its polar vortex array configuration, albeit with a net $+P_x$ polarization (Fig.~\ref{PE_loop_100K}f-g). Overall, the $P_x$-$E_x$ hysteresis loop in the \PTOSTO~superlattice, resembling that of a ferroelectric, is considerably slimmer than that in a single-crystal, single-domain PbTiO$_3$ (see the blue loop in Fig.~\ref{PE_loop_100K}a). It is notable that the superlattice has a much lower switching field, likely due to the high-density of 90$^\circ$ domain walls within the vortex arrays. Our MD simulations are consistent with recent experiments where the
(STO)$_{20}$/(PTO)$_{20}$/(STO)$_{20}$ trilayer grown on the DyScO$_3$ substrate showed a reduced in-plane coercive field of $\approx$0.03 MV/cm~\cite{Behera23p2208367}. 

\begin{figure}[]
\centering
\includegraphics[width=1.0\textwidth]{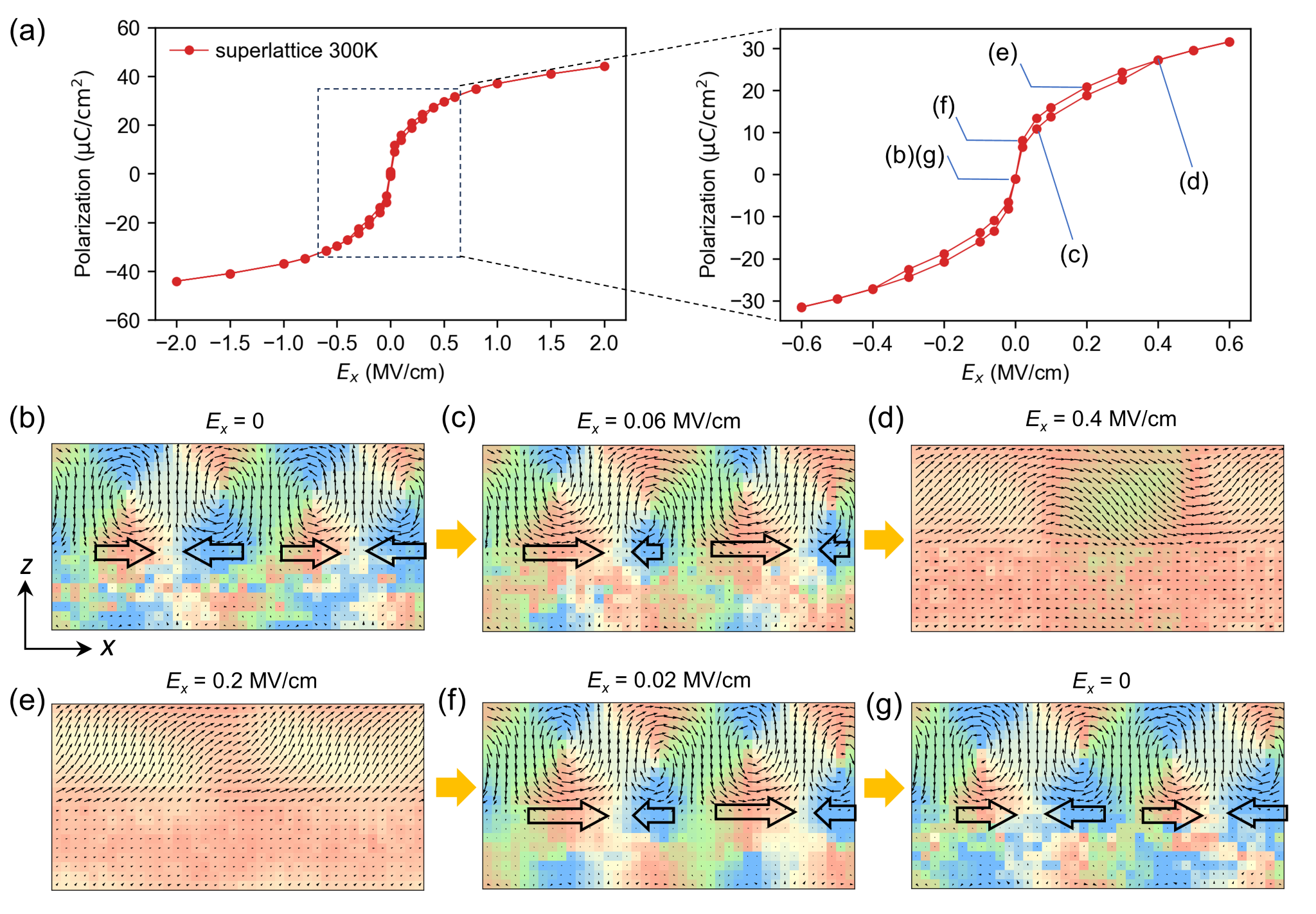}
\caption{Polarization-electric field ($P_x$-$E_x$) hysteresis loop of the AFE$^*$ state featuring aligned vortex cores in the \PTOSTO~superlattice at 300 K. The right figure shows a magnified view of the loop for $E_x$ in the range between $-0.6$ MV/cm and 0.6 MV/cm. 
Snapshots of topological structures throughout the loop, obtained from MD simulations, are presented in (b)-(g), using the color scheme from Fig.~1.
}
\label{PE_loop_300K}
\end{figure}

In contrast, the simulated $P_x$-$E_x$ loop at 300~K shows no remnant polarization and is nearly hysteresis-free (Fig.~\ref{PE_loop_300K}a). The S-shaped curve differs from the double hysteresis loop typically associated with antiferroelectric materials~\cite{Cao19p65,Toledano16p014107}. Instead, it bears a resemblance to the behavior seen in ``superparaelectric" systems~\cite{Chu21p33,Pan21p100}. 
Previously, Aramberri~\etal~employed a second-principles method to simulate (PTO)$_4$/(STO)$_4$ superlattices and also observed the evolution from a double hysteresis loop at 0~K to a nonhysteric S-shaped curve at room temperature~\cite{Aramberri22peabn4880}.
As illustrated in Fig.~\ref{PE_loop_300K}b, the initial configuration features $+P_x$ and $-P_x$ domains of equal volume with aligned vortex cores. The application of $+E_x$ drives opposing movements of neighboring vortex cores. This process is coupled with the growth of $P_x$ domains (Fig.~\ref{PE_loop_300K}c). At a sufficiently high field of 0.4 MV/cm, the vortex state evolves smoothly to a dipole wave state (Fig.~\ref{PE_loop_300K}d) and eventually to a single-domain state with $+P_x$. As the field is reduced, the superlattice undergoes successive transitions from an $a$-domain to dipole waves and back to vortex arrays in the PTO layers (Fig.~\ref{PE_loop_300K}f).  At zero field, polar vortex arrays return to the original state without net in-plane polarization (Fig.~\ref{PE_loop_300K}g). The spontaneous formation of polar vortex lattices indicates a one-to-one correspondence between the topological structure and the mechanical and electrical boundary conditions, which include factors such as epitaxial strain and lattice periodicity, within a superlattice at a given temperature. These conditions collectively determine the energetically favorable polar configuration and act as a restorative force, reforming vortex arrays upon field removal. This explains the hysteresis-free, superparaelectric-like $P_x$-$E_x$ curve.

\subsection{In-plane dielectric susceptibility and tunability}

\begin{figure}[]
\centering
\includegraphics[width=0.6\textwidth]{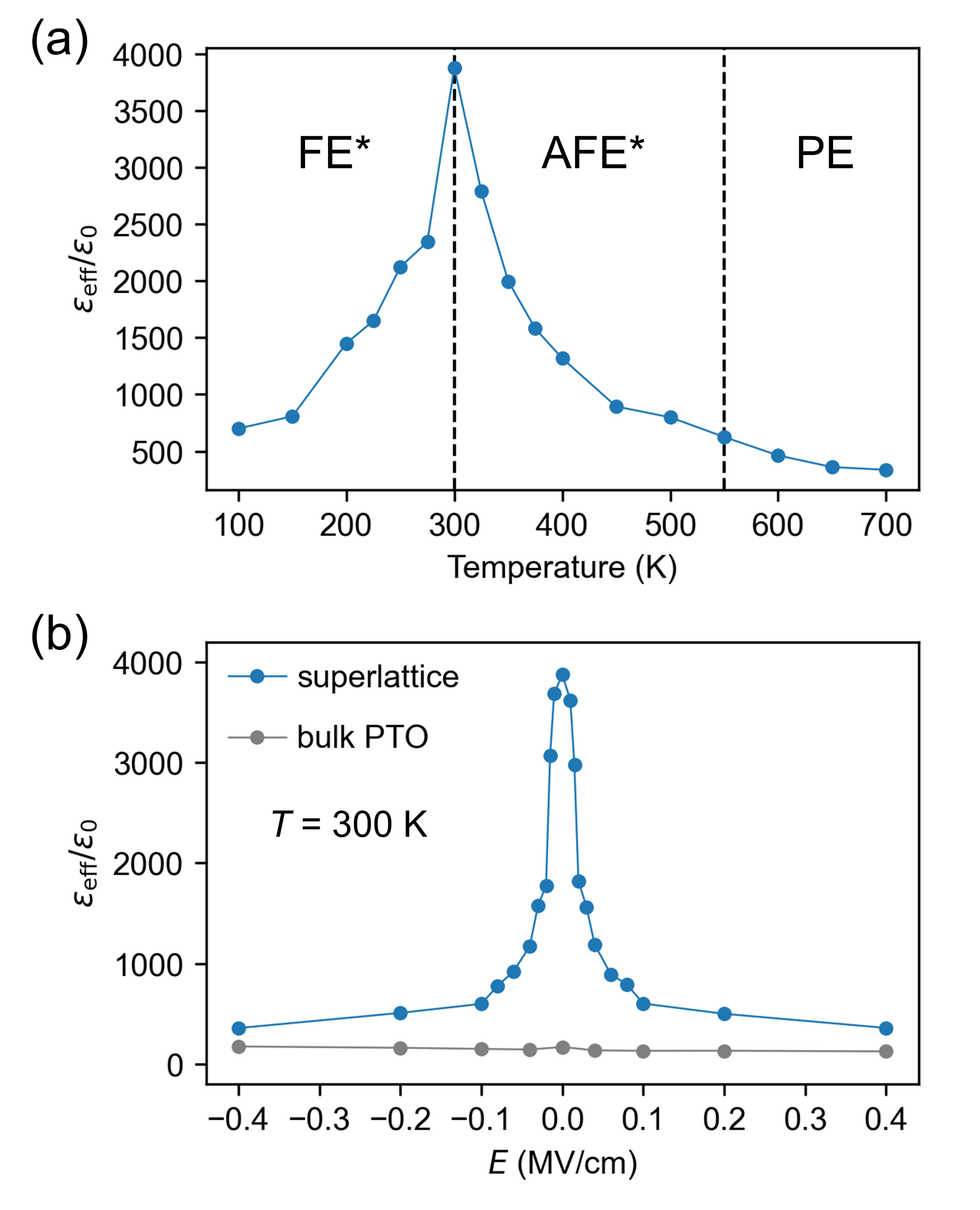}
\caption{ 
(a) Temperature dependence of the in-plane dielectric constant (in units of vacuum permittivity $\epsilon_0$) in the \PTOSTO~superlattice. (b) Field-dependent in-plane dielectric constant of the superlattice supporting polar vortex arrays at 300~K, compared to that of the bulk PTO at the same temperature.
}
\label{dielectric}
\end{figure}

We extract the temperature dependence of the in-plane static dielectric constant $\epsilon_\mathrm{eff}$ from the $P_x$-$E_x$ loops at various temperatures. For a typical ferroelectric, the dielectric constant often undergoes a discontinuous change at the FE-PE phase transition~\cite{Samara71p277}. Here, we find that the in-plane dielectric constant of the superlattice also exhibits a discontinuous change at the FE$^*$-AFE$^*$ transition (Fig.~\ref{dielectric}a). In contrast, the dielectric constant remains nearly continuous at the AFE$^*$-PE transition, likely due to the intermediate polar liquid phase (Fig.~\ref{polar_liquid}) that smooths the dielectric response across the transition temperature. Previous studies have demonstrated that the position of the vortex core can be modulated by strain~\cite{Gong21p5503,Yang21pL220303}, suggesting that the FE$^*$-AFE$^*$ transition temperature (or the peak of the dielectric constant) can be tuned to accommodate different application requirements. 

We compare the in-plane dielectric constant of the superlattice and the out-of-plane dielectric constant of PTO under varying electric fields at 300~K. 
For a given electric field, we first obtained the equilibrium domain structure with finite-field MD simulations. 
Subsequently, we applied a much smaller in-plane probing electric field of approximately 0.02 MV/cm (in addition to the background field) to perturb the polarization. The in-plane dielectric constant is then computed from the slope of the polarization-electric field curve. The probing electric field for susceptibility measurements is significantly smaller than the background electric field to ensure the validity of linear response theory.
As plotted in Fig.~\ref{dielectric}b, our results indicate that the dielectric constant of the superlattice is highly tunable, reaching a maximum value of 3900 in units of $\epsilon_0$ (vacuum permittivity), while the dielectric constant of PTO is only 170, consistent with experimental values reported in the literature~\cite{Li93p313,Ikegami71p1060}. This significant enhancement in the in-plane dielectric constant of the superlattice, compared to bulk PTO, underscores the potential of these topological structures for applications requiring high dielectric susceptibility and tunability.

\subsection{Pb-doping-driven dipolar topological transitions in (PTO)$_{10}$/(PSTO)$_{10}$ superlattices}
\begin{figure}[]
\centering
\includegraphics[width=0.8\textwidth]{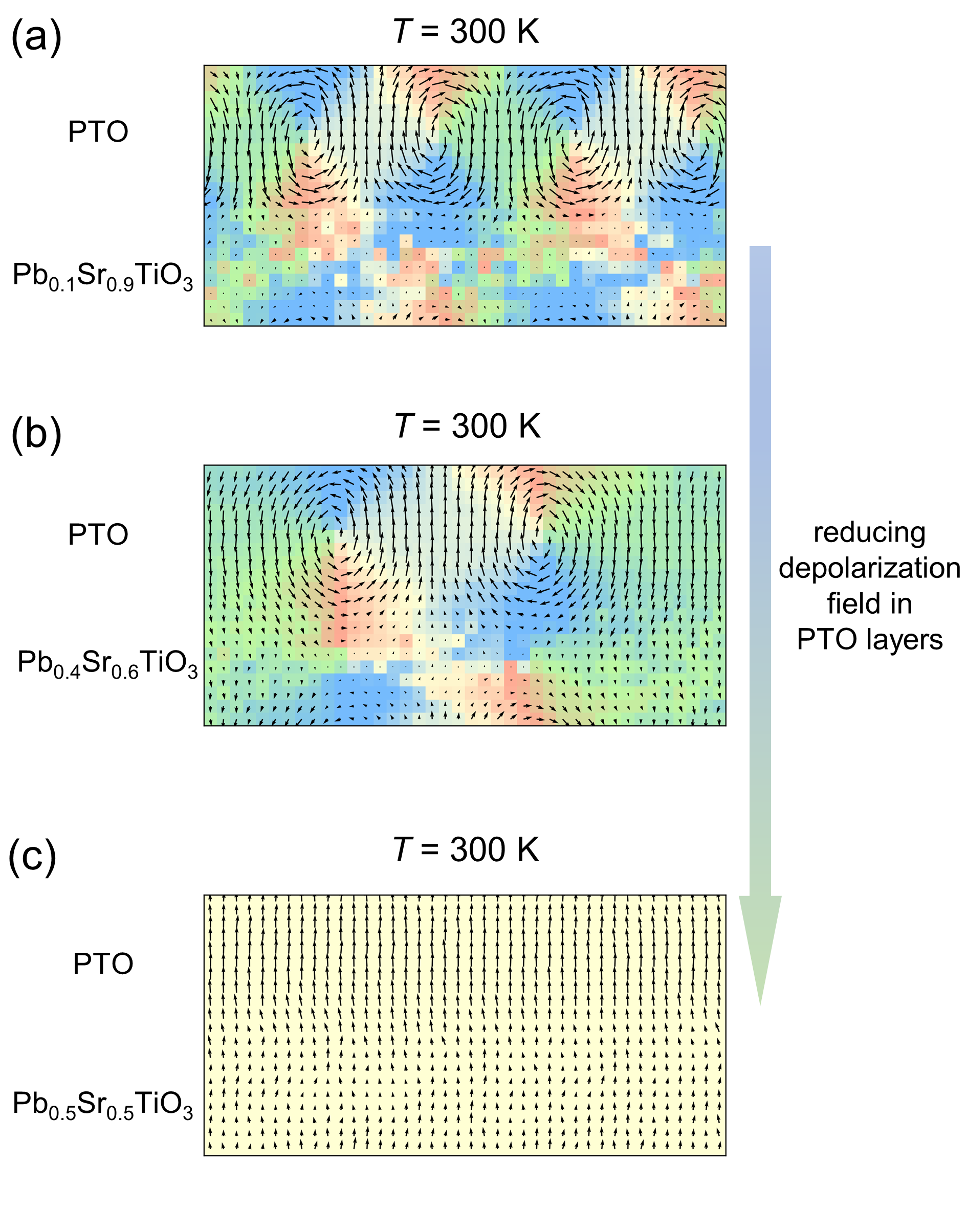}
\caption{Pb-doping-induced dipolar topological transitions in (PTO)$_{10}$/(PSTO)$_{10}$ superlattices as viewed in the $xz$ plane.
Cross-sectional views show that as the Pb concentration in the STO layers increases, the number of vortex pairs in the PTO layer decreases from (a) two pairs to (b) one pair, ultimately transforming into a single domain in (c). The color scheme is identical to that used in Fig.~1.
}
\label{doped_structure}
\end{figure}

The impact of doping on topological textures in superlattices, along with their electrical and thermal responses, remains underexplored. Here, we investigate the main effect of Pb doping in the STO layers: the weakening of the depolarization field in the adjacent PTO layers. 
As the Pb doping concentration ($\alpha$) increases in the STO layers (Pb$_\alpha$Sr$_{1-\alpha}$TiO$_3$, refereed to as PSTO), we observe a decrease in the vortex density within the PTO layers in the $xz$ plane. This reduction is attributed to the widening of $P_z$ domains, as evident when comparing $\alpha = 0.1$ in Fig.~\ref{doped_structure}a and $\alpha = 0.4$ in Fig.~\ref{doped_structure}b. Although the topological structure in the PTO layer resembles a vortex lattice in \PTOPSTO~with $\alpha = 0.4$,
further analysis of the three-dimensional dipole distributions reveals the formation of enlarged skyrmion bubbles embedded within $c$ domains (see discussions of Fig.~\ref{doped_phase_transition}). At a critical doping concentration of $\alpha=0.5$, the topological configurations vanish entirely, and the system transitions into a single-domain state, as depicted in Fig.~\ref{doped_structure}c. 

The abrupt increase in domain width associated with enhanced Pb doping in superlattices can be understood using the theoretical framework developed by Bratkovsky and Levanyuk~\cite{Bratkovsky00p3177}. Their model describes the domain width in ferroelectric thin films as a function of the thickness of a nonferroelectric ``dead" or ``passive" layer at the film-electrode interface. This layer's thickness governs the strength of the depolarization field. In our study, Pb doping in the STO layers effectively mimics the role of this passive interfacial layer, with the doping concentration regulating the depolarization field strength. 
As the domain width decreases, the number of domain walls increases, elevating polarization gradient energy. Conversely, a smaller domain width reduces the energy of polarization-depolarization field coupling. The interplay between these two main energy terms determines that the domain width in ferroelectric thin films increases exponentially as the depolarization field weakens, eventually tending towards infinity, or reaching the single-domain limit. This trend aligns well with our observations in \PTOPSTO~superlattices, where the Pb doping in the STO layers effectively weakens the depolarization field.

%Our pattern here still conforms to this theory, indicating that the theory still holds at the nanoscale, but with unique manifestations, likely accompanied by the evolution of topological structures. 

\subsection{Temperature-driven phase transitions in (PTO)$_{10}$/(PSTO)$_{10}$  superlattices}
\begin{figure}[]
\centering
\includegraphics[width=0.85\textwidth]{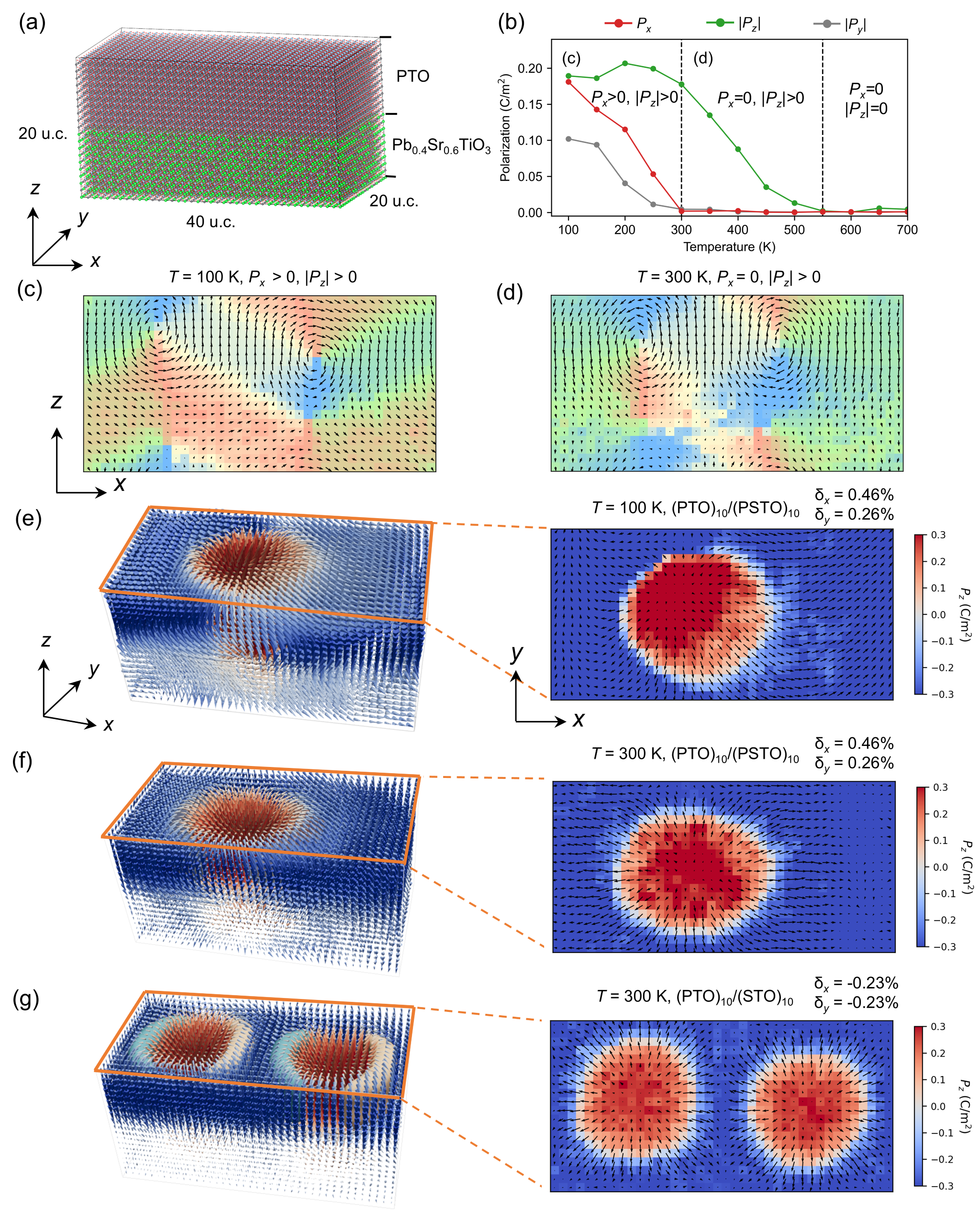}
\caption{(a) A 40$\times$20$\times$20 supercell used to model the (PTO)$_{10}$/(Pb$_{0.4}$Sr$_{0.6}$TiO$_3$)$_{10}$ superlattice with periodic boundary conditions. (b) Temperature dependence of the average polarization of the superlattice along the $x$, $y$ and $z$ axes. Cross-sectional views of the enlarged skyrmion bubble in the $xz$ plane at (c) 100 K and (d) 300 K, using the color scheme from Fig.~1.
Three-dimensional dipole distributions of diffusive skyrmion bubbles in  the (PTO)$_{10}$/(Pb$_{0.4}$Sr$_{0.6}$TiO$_3$)$_{10}$ superlattice at (e) 100~K and (f) 300~K are displayed on the left, with $xy$ cross-sections shown on the right. 
For comparison, dipole distributions of standard skyrmion bubbles in the (PTO)$_{10}$/(STO)$_{10}$ superlattice are presented in (g). The background colors in (e-g)
represent local $P_z$ values.}
\label{doped_phase_transition}
\end{figure}

We select a representative composition of PSTO with $\alpha=0.4$ to explore the temperature-dependent phase transitions in (PTO)$_{10}$/(PSTO)$_{10}$ superlattices, using 40$\times$20$\times$20 supercells at a strain state of $\delta_{x}=0.46$\% and $\delta_{y}=0.26$\% (Fig.~\ref{doped_phase_transition}a). The Pb doping recovers  the out-of-plane polarization ($P_z$), which generally decreases with increasing temperature. The system undergoes a conventional ferroelectric-paraelectric phase transition along the out-of-plane direction at $T\approx 550$~K (see $P_z$ versus $T$ in Fig.~\ref{doped_phase_transition}b). Similar to the FE$^*$-AFE$^*$-PE successive phase transitions observed in the \PTOSTO~superlattice along the $x$ axis, the \PTOPSTO~superlattice maintains this transition sequence (see $P_x$ versus $T$ in Fig.~\ref{doped_phase_transition}b). As shown in Fig.~\ref{doped_phase_transition}c, the zigzag ordering of vortex cores in the $xz$ plane at 100~K results in a net $P_x$, which evolves to a horizontally aligned state at 300~K (Fig.~\ref{doped_phase_transition}d) associated with zero $P_x$ components.
We note that although the spacial distributions of dipoles in the $xz$ plane resemble polar vortex arrays at both 100~K and 300~K, detailed three-dimensional dipole distribution analysis confirms that the dipoles form enlarged skyrmion bubbles embedded in $c$ domains (see Fig.~\ref{doped_phase_transition}e-f). 

By comparing with standard skyrmion bubble formed in undoped \PTOSTO~grown on an STO substrate ($\delta_{x}=$ -0.23~\%, $\delta_{y}=$ -0.23~\%), we identify several distinct features of the enlarged skyrmion bubbles. The dipole distributions in the $xy$ plane of standard skyrmion bubbles (Fig.~\ref{doped_phase_transition}g) 
reveal sharp boundaries and a low volume fraction of pure $c$ domains. In contrast, at 100~K, the skyrmion bubble residing in $c$ domains features diffusive boundaries (Fig.~\ref{doped_phase_transition}e). The bubble is anisotropic, consistent with the non-zero $P_y$ component shown in Fig.~\ref{doped_phase_transition}b. 
At 300~K, the enlarged bubble becomes more isotropic (Fig.~\ref{doped_phase_transition}f), associated with a zero net $P_y$.

\subsection{Polarization-electric field hystersis loops in (PTO)$_{10}$/(PSTO)$_{10}$  superlattices}

\begin{figure}[]
\centering
\includegraphics[width=1.0\textwidth]{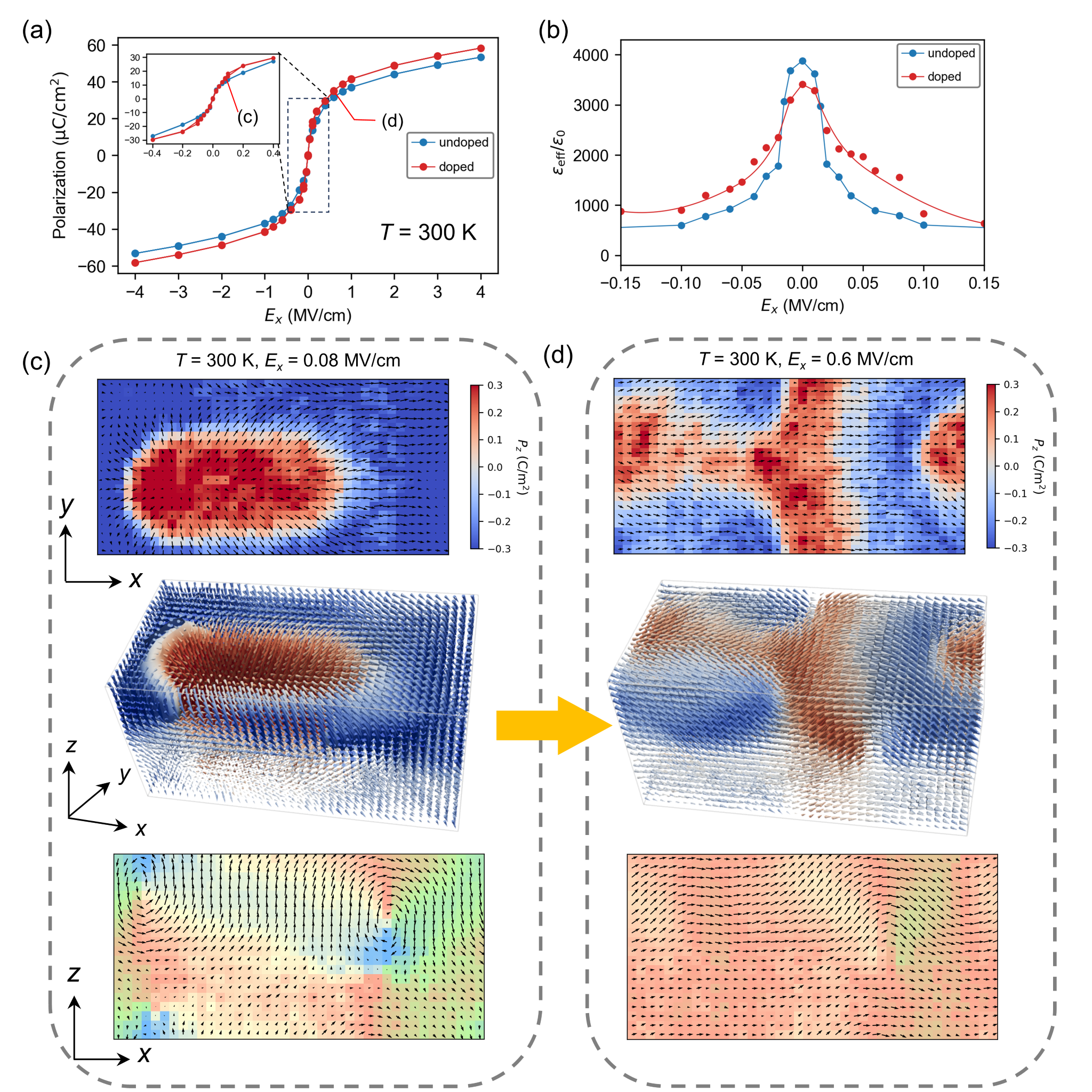}
\caption{Comparison of (a) $P_x$-$E_x$ loops and (b) dielectric properties for doped ($\alpha=0.4$) and undoped superlattices. Dipole distributions in the \PTOPSTO~superlattice at 300~K are shown for (c) $E_x=0.08$~MV/cm and (d) $E_x=0.6$~MV/cm. The color scheme is the same as that in Fig.~8.
}
\label{Efield doped}
\end{figure}

As shown in Fig.~\ref{Efield doped}a, the evolution of $P_x$ in response to $E_x$ in \PTOPSTO~at 300~K adopts an S-shaped loop that is nearly lossless, similar to that observed in \PTOSTO~at the same temperature. Additionally,  
we find that the doped superlattice becomes more susceptible to the electric field when $E_x>0.02$~MV/cm, where the dielectric constant of the \PTOPSTO~superlattice becomes substantially higher than that of the \PTOSTO~superlattice (Fig.~\ref{Efield doped}b).

Although the $P_x$-$E_x$ loop of the doped superlattice resembles that of the undoped \PTOSTO~superlattice, the evolution of the topological structures is notably different. 
Specifically, at an electric field strength of 0.08 MV/cm, the skyrmion bubble becomes elongated along the $x$ axis (see top and middle panels in Fig.~\ref{Efield doped}c). In the $xz$ plane, this elongation is manifested by the opposing movements of neighboring vortex cores along the $x$ axis (see bottom panel in Fig.~\ref{Efield doped}c). When the field strength exceeds 0.6 MV/cm, the system transitions into a complex 3D dipole wave (Fig.~\ref{Efield doped}d). As expected, the superlattice dopts a single-domain state with $+P_x$ only when $E_x>2$ MV/cm. 

\subsection{Concerted effects of doping, strain, and lattice periodicity}

\begin{figure}[]
\centering
\includegraphics[width=1.0\textwidth]{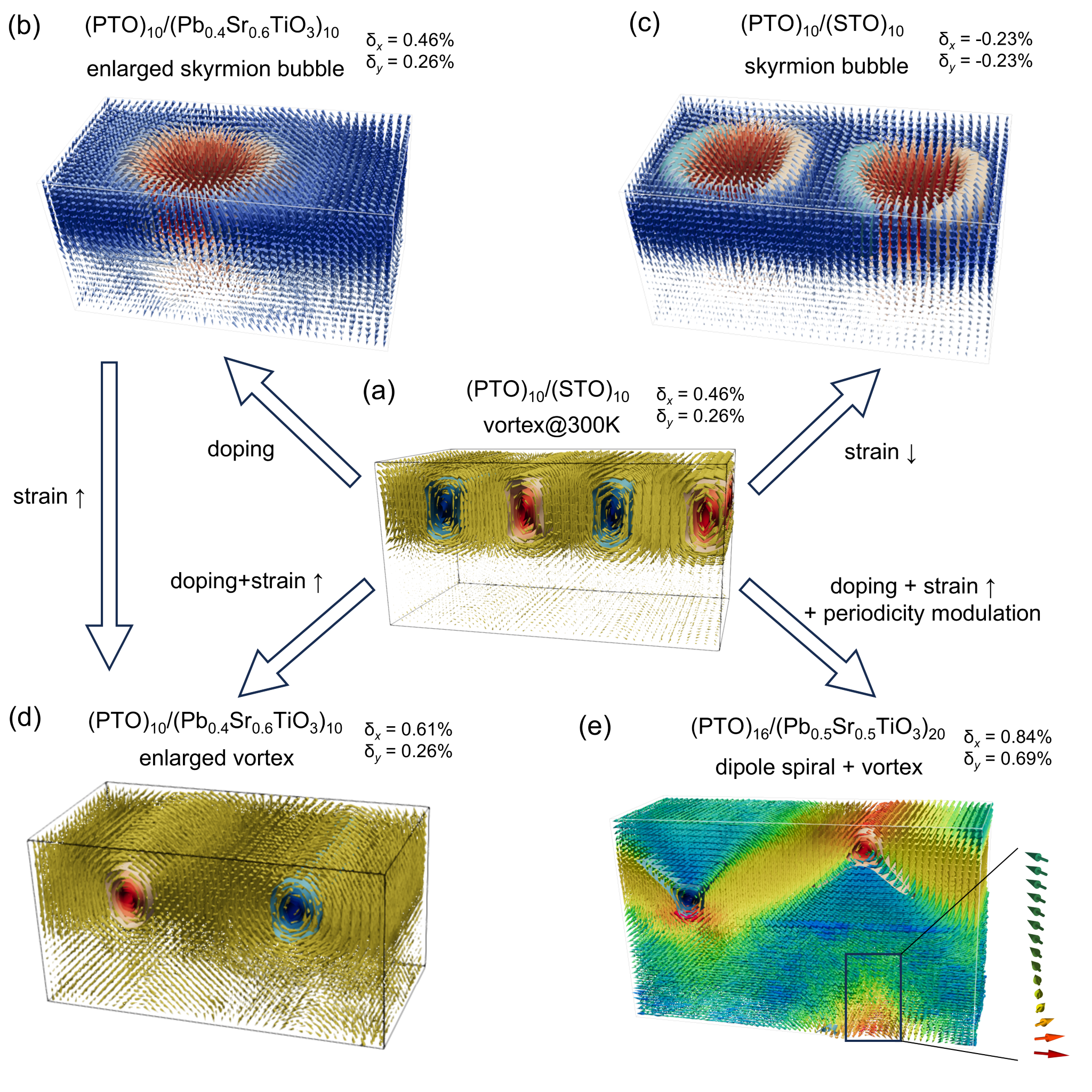}
\caption{
Concerted effects of doping, strain, and lattice periodicity on the dipolar topologies. MD simulations with different electrical and mechancial boundary conditions at 300~K reveal (a) polar vortex arrays, (b) enlarged skyrmion bubbles in $c$ domains, (c) skyrmion bubbles, (d) enlarged  vortex arrays, and (e) dipole spirals connecting vortices~\cite{Hu24p046802} in different perovskite superlattices.
}
\label{strain+doping}
\end{figure}

Doping strategies can also be combined with epitaxial strain engineering and superlattice periodicity control to realize diverse polar topological structures. 
Previously, it has been established experimentally that epitaxial strain engineering can be employed to achieve skyrmion bubbles in (PTO)$_{16}$/(STO)$_{16}$ grown on a STO substrate with a slightly compressive strain~\cite{Das19p368}, as well as polar vortex arrays in \PTOSTO~grown on a DyScO$_3$ substrate with a tensile strain~\cite{Yadav16p198}. This trend is reproduced by our MD simulations here (Fig.~~\ref{strain+doping}a-c). As discussed above, the introduction of Pb dopants into the STO layers at $\delta_x=0.46$\% and $\delta_y=0.26$\% can drive the superlattices from the vortex state to a mixed state with enlarged skyrmion bubbles and $c$ domains (Fig.~\ref{strain+doping}b). Interestingly, further applying an in-plane tensile strain to $\delta_x=0.61$\% and $\delta_y=0.26$\% will recover the vortex arrays featuring a larger distance between vortex cores, as shown in Fig.~\ref{strain+doping}d. An even more complex topology can be achieved by simultaneously adjusting the strain, doping concentration, and the lattice periodicity. For example, as recently demonstrated in our work,  the (PTO)$_{16}$/(Pb$_{0.5}$Sr$_{0.5}$TiO$_3$)$_{20}$ superlattice at $\delta_x=0.84$\% and $\delta_y=0.69$\% supports vortex arrays in PTO layers while forming dipole spirals in the Pb$_{0.5}$Sr$_{0.5}$TiO$_3$ layers~\cite{Hu24p046802}. Large-scale MD simulations at finite temperatures, enabled by accurate deep potential capable of describing superlattices of different compositions, thus offer exciting opportunities to discover new types of ferroelectric topological structures and their dynamical properties. 

\section{Conclusions}
Our large-scale molecular dynamical simulations, enabled by accurate machine-learning force fields, demonstrate that the topologically engineered \PTOSTOall~superlattice exhibits a range of unique characteristics along the in-plane direction. These include a low coercive field, distinctive phase transitions, and promising potential for tunable dielectric applications. A particularly intriguing finding is the counterintuitive temperature-driven transition from a ferroelectric-like (ferrielectric-like) to an antiferroelectric-like state. This unexpected behavior underscores the potential of harnessing topological domain structures to engineer properties that are unattainable in conventional single-domain states.
The regulation of the depolarization field in the PbTiO$_3$ layers through strategic doping of the SrTiO$_3$ layers is effective in driving dipolar topological phase transitions and fostering the formation of new types of complex polar topologies.  The quantitative atomistic insights into the thermal and electrical responses of these perovskite superlattices provide a foundation for tailoring their properties for nanoelectronic applications.

\begin{acknowledgments}
This work is supported by National Natural Science Foundation of China (92370104) and Natural Science Foundation of Zhejiang Province (2022XHSJJ006). The computational resource is provided by Westlake HPC Center.
\end{acknowledgments}

\bibliography{SL}

\clearpage

\begin{table}[h]
\caption{BEC tensors for ions in PTO and STO. $Z_{11}^*$, $Z_{22}^*$, and $Z_{33}^*$ denote the diagonal components of the tensor. The BEC tensors for ions in Pb$_\alpha$Sr$_{1-\alpha}$TiO$_3$ are obtained from the linear interpolation of the BEC tensors in PTO and STO. For example, for Pb$_{0.4}$Sr$_{0.6}$TiO$_3$, $Z_\mathrm{A}^* = 0.4Z_\mathrm{Pb}^* + 0.6Z_\mathrm{Sr}^*$.}
\centering
\begin{tabularx}{\textwidth}{>{\centering\arraybackslash}p{3cm}|*{3}{>{\centering\arraybackslash}X}}
\hline
\multirow{2}{*}{\centering System} & $Z_\mathrm{A}^*$ (e) & $Z_\mathrm{Ti}^*$ (e) & $Z_\mathrm{O}^*$(e) \\
\cline{2-4}
        & $Z_{11}^*$ \hspace{0.4em} $Z_{22}^*$ \hspace{0.4em} $Z_{33}^*$ & $Z_{11}^*$ \hspace{0.4em} $Z_{22}^*$ \hspace{0.4em} $Z_{33}^*$ & $Z_{11}^*$ \hspace{0.4em} $Z_{22}^*$ \hspace{0.4em} $Z_{33}^*$ \\
\hline
PTO & 3.74 \hspace{0.4em} 3.74 \hspace{0.4em} 3.45 & 6.17 \hspace{0.4em} 6.17 \hspace{0.4em} 5.21 & -3.30 \hspace{0.4em} -3.30 \hspace{0.4em} -2.89 \\
\hline
STO & 2.56 \hspace{0.4em} 2.56 \hspace{0.4em} 2.56 & 7.40 \hspace{0.4em} 7.40 \hspace{0.4em} 7.40 & -3.32 \hspace{0.4em} -3.32 \hspace{0.4em} -3.32 \\
\hline
Pb$_{0.4}$Sr$_{0.6}$TiO$_3$ & 3.03 \hspace{0.4em} 3.03 \hspace{0.4em} 2.92 & 6.91 \hspace{0.4em} 6.91 \hspace{0.4em} 6.52 & -3.31 \hspace{0.4em} -3.31 \hspace{0.4em} -3.15 \\
\hline
\end{tabularx}
\label{BECs}
\end{table}

\clearpage
\newpage

\clearpage
\newpage

\clearpage
\newpage

\clearpage
\newpage

\clearpage
\newpage

\clearpage
\newpage

\clearpage
\newpage

\clearpage
\newpage

\clearpage
\newpage

\clearpage
\newpage

\end{document}